\documentclass{emulateapj}

\newcommand{\OII}{[O {\sc ii}]}

\begin{document}

\title{ENVIRONMENTAL EFFECTS ON THE STAR FORMATION ACTIVITY IN GALAXIES
       AT $z \simeq 1.2$ IN THE COSMOS FIELD\altaffilmark{*}}

\author{Y. Ideue      \altaffilmark{1},
        T. Nagao      \altaffilmark{1,2},
        Y. Taniguchi  \altaffilmark{2},
	Y. Shioya     \altaffilmark{2},
	T. Saito      \altaffilmark{2}, 
        T. Murayama   \altaffilmark{3},
	S. Sasaki     \altaffilmark{1,3},
        J. R. Trump   \altaffilmark{4},
        A. M. Koekemoer\altaffilmark{5},
        H. Aussel     \altaffilmark{6},
        P. Capak      \altaffilmark{7},
        O. Ilbert     \altaffilmark{8},
        H. McCracken  \altaffilmark{9},
        B. Mobasher   \altaffilmark{10},
        M. Salvato    \altaffilmark{7},
        D. B. Sanders    \altaffilmark{8}, and
        N. Scoville   \altaffilmark{7}
        }

\altaffiltext{*}{Based on observations with the NASA/ESA 
        {\it Hubble Space Telescope}, obtained at the Space Telescope Science 
	Institute, which is operated by AURA Inc, under NASA contract NAS 
	5-26555. Also based on observations made with the Spitzer Space 
	Telescope, which is operated by the 
	Jet Propulsion Laboratry, California Institute of Technology, 
	under NASA contract 1407. Also based on data collected at;  
	the Subaru Telescope, which is operated by the National Astronomical 
	Observatry of Japan; the XMM-Newton, an ESA science mission with 
	instruments and contributions directly funded by ESA Member States and
	NASA; the European Southern Observatory under Large 
	Program 175.A-0839, Chile; Kitt Peak National Observatory, Cerro 
	Tololo Inter-American Observatory and the National Optical Astronomy 
	Observatory, which are operated by the Association of Universities for 
	Research in Astronomy, Inc. (AURA) under cooperative agreement with 
	the National Science Foundation; and the Canada-France-Hawaii 
	Telescope with MegaPrime/MegaCam operated as a joint project by the 
	CFHT Corporation, CEA/DAPNIA, the NRC and CADC of Canada, the CNRS 
	of France, TERAPIX and the Univ. of Hawaii.}

\altaffiltext{1}{Graduate School of Science and Engineering, Ehime University, 
        Bunkyo-cho, Matsuyama 790-8577, Japan}

\altaffiltext{2}{Research Center for Space and Cosmic Evolution, 
        Ehime University, Bunkyo-cho, Matsuyama 790-8577, Japan}

\altaffiltext{3}{Astronomical Institute, Graduate School of Science,
        Tohoku University, Aramaki, Aoba, Sendai 980-8578, Japan}

\altaffiltext{4}{Steward Observatory, University of Arizona, Tucson, AZ 85721}

\altaffiltext{5}{Space Telescope Science Institute, 3700 San Martin Drive, Baltimore, MD 21218}

\altaffiltext{6}{AIM Unit\'e Mixte de Recherche CEA CNRS Universit\'e Paris VII
UMR n158 
        }

\altaffiltext{7}{California Institute of Technology, MC 105-24, 
        1200 East California Boulevard, Pasadena, CA 91125}

\altaffiltext{8}{Institute for Astronomy, University of Hawaii, 
        Honolulu, HI 96822}

\altaffiltext{9}{Institut d'Astrophysique de Paris, UMR 7095, CNRS,
        Universit Pierre et Marie Curie, 98 bis Boukevard Arago, 
        F-75014 Paris, France.}

\altaffiltext{10}{Department of Physics and Astronomy, University of 
        California, Riverside, CA 92521, USA}

\shortauthors{Ideue et al.}
\shorttitle{The star formation activity \- density relation at $z \simeq 1.2$}

\begin{abstract}
We investigate the relation between the star-formation activity in galaxies 
and environment at $z\simeq 1.2$ in the COSMOS field, using the fraction of 
\OII\ emitters and the local galaxy density.
The fraction of \OII\ emitters appears to be almost constant over the surface
density of galaxies between 0.2 and 10 Mpc$^{-2}$.
This trend is different from that seen in the local universe 
where the star-formation activity is weaker in higher density regions. 
To understand this difference between $z\sim1$ and $z\sim 0$, we study the
fraction of non-isolated galaxies as a function of local galaxy density.
We find that the fraction of non-isolated galaxies increases with increasing
density. Our results suggest that the star 
formation in galaxies at $z \sim 1$ is triggered by galaxy 
interaction and/or mergers.

\end{abstract}



\section{INTRODUCTION}
 
To understand the formation and evolution of galaxies strongly relies on 
investigations of when and where stars formed in the history of the universe 
from high redshift to the present day. 
Since the pioneering study by Madau et al. (1996), a number of observational 
studies have investigated how the star formation rate density (SFRD) varies as
a function of redshift. 
Such studies brought us a global evolutionary picture of the SFRD; 
the SFRD steeply increases in the first 900 Myrs, peaks at $z\sim 3-1$, 
and then decreases rapidly toward the present day 
(e.g., Madau et al.1996; Lilly et al. 1996; Fujita et al. 2003; Bouwens \& 
Illingworth 2006). 
However, we are still far from a complete understanding of the galaxy 
evolution since the redshift evolution of the SFRD gives us only the 
``averaged'' picture of the star formation history in the universe. 
There exists a wide variety of environments in the universe, such as field, 
groups, and clusters of galaxies. Moreover, even within the field, there are structures 
like high- and low-density regions.
It is thus essential to study the star formation in galaxies as a function of 
both the redshift and the environment.
One possible approach is to study the relation between the galaxy properties
and the environments at various epochs.   

The environmental effects on galaxy properties in the local universe have been
extensively studied in the past. Star formation activity in galaxies is known
to decline with increasing density (e.g., Lewis et al. 2002; Gomez et al.2003),
and the fraction of star-formation galaxies decreases with increasing density
(e.g., Carter et al. 2001; Balogh et al. 2004; Mateus \& Sodr\'e 2004).
These findings agrees quite well with the so-called 
``morphology-density relation'', i.e., the fraction of early-type galaxies
increases with increasing density, while the fraction of late-type galaxies
decreases with increasing density (e.g., Dressler 1980; Dressler et al. 1997; 
Goto et al. 2003; Capak et al. 2007). All these results suggests that the star
formation in high-density region are more active than in the low-density
regions. In contrast, recent observational studies in the Great Observatories
Origins Deep Survey (GOODS; Dickinson \& Giavalisco 2002) and the DEEP2 Galaxy
Redshift Survey (DEEP2; Davis et al. 2003) shows that this relation becomes
completely opposite at $z\sim 1$; the star-formation activity depends 
positively on the density of galaxies (Elbaz et al. 2007; Cooper et al. 2008).
The origin of this drastic change between $z\sim 1$ and $z\sim 0$ in still
not clear.

In this paper, we present the relationship between the star-formation activity 
and the environment at $z \simeq 1.2$, in the Cosmic Evolution Survey 
(COSMOS; Scoville et al. 2007a) field. 
We estimate star-formation activity by focusing on the ``emitter fraction''.
Specifically we investigate the fraction of galaxies emitting \OII\ 
$\lambda$3727 emission (hereafter ``\OII\ emitters''), as a function of the 
galaxy local density. The \OII\ $\lambda 3727$ (hereafter \OII\ ) emission line 
is a good estimator of the star-formation activity in galaxies at intermediate 
redshift (e.g., Jansen et al. 2001).
Although it is known that the \OII\ emission line is affected by dust 
extinction and metallicity, such effects on the emitter fraction are thought
to be small, since our [OII] emitters are selected based on the equivalent
widths (EWs).
Specifically, the \OII\ EW is significantly smaller in very metal-poor and
metal-rich galaxies, with respect to galaxies with the solar metallicity. 
However, it is observed that the variation is relatively small, only a factor 
of 2-3, in the metallicity range of $7.7<12+{\rm log (O/H)}<9.2$ 
(e.g., Nagao et al. 2006). 
Since galaxies with metallicity out of this range are expected to be rare, 
we consider that the metallicity effect on the emitter fraction is not 
significant.

We have already carried out the Subaru imaging observations of the COSMOS 
field (Taniguchi et al. 2007).
The sample of \OII\ emitters in the COSMOS field was obtained by Takahashi 
et al. (2007) using a narrowband filter NB816 
[$\lambda_c = 8150 $ \AA\ and $\Delta \lambda {\rm (FWHM)} = 120 $ \AA\ ]
(see also Murayama et al. 2007 and Shioya et al. 2008).
Thanks to the very large survey area (2 square degrees) of the COSMOS field,  
we are able to obtain a typical picture of the star-formation activity in 
galaxies, avoiding the cosmic variance. 
The COSMOS field includes various regions within a wide of density
(e.g., Scoville et al. 2007a, 2007b; Takahashi et al. 2007), so that we 
are able to investigate systematically the environmental effects on the star 
formation activity in galaxies.

In \S 2, we describe the sample selection and discuss the possible 
contamination in the photometric-redshift sample. 
The results of the star-formation activity-environment relation at 
$z\simeq 1.2$ and the dependence of the morphology on the density are 
presented in \S 3. The interpretation of our results is discussed in \S 4.
And finally we give a brief summary in \S 5.
Throughout this paper, magnitudes are given in the AB system.
We adopt a flat universe with the following cosmological parameters; 
$\Omega_{\rm matter} = 0.3$, $\Omega_{\Lambda} = 0.7$,
and $H_0 = 70\; {\rm km\;s^{-1}\;Mpc^{-1}}$. 

\section{THE SAMPLE}

\subsection{Sample Selection}

In order to derive the emitter fraction and investigate its environmental
dependence, we require two samples of galaxies;
the \OII\ emitter sample and the 
total galaxy sample (i.e., both galaxies with and without \OII\ emission).
The former one is used to identify actively star-forming galaxies. The latter 
one is necessary to estimate the galaxy local density for each position in the
COSMOS field.

We carried out the imaging survey for \OII\ emitters at 
$z\simeq 1.2$ with a narrow-band filter NB816 (see Ajiki et al. 2003 for 
details) in the COSMOS 2 square degree field, using the Suprime-Cam (Miyazaki 
et al. 2002) on the Subaru Telescope (Kaifu et al. 2000; Iye et al. 2004). 
This survey covers a sky area of 6700 arcmim$^2$ in the COSMOS field, and a 
redshift range between 1.17 and 1.20 for \OII\ 
($\Delta z=0.03$, corresponds to 70 Mpc in the comoving distance), 
corresponding to a survey volume of $5.5\times 10^5~$Mpc$^3$.
We have obtained a sample of 3176 \OII\ emitters from this data, down to 
$i'\simeq 26.1$. The selection threshold of the \OII\ EW is $>12$ \AA\ in the
observed frame, or $>12$ \AA\ in the rest frame 
(see Takahashi et al. 2007 for more details). We use a sub-sample of this 
parent sample of \OII\ emitters for our analysis and discussion.

To identify both galaxies with and without \OII\ emission in the range of 
$1.17 < z < 1.20$ (i.e., the coverage of the NB816 filter for \OII\ emitters), 
we select galaxies with a photometric redshift (photo-$z$, $z_{\rm ph}$) in the
range of $1.17< z_{\rm ph} < 1.20$ from the COSMOS photo-$z$ catalog 
(Ilbert et al. 2008). 
Since fainter galaxies have larger photo-$z$ errors, we select only bright 
galaxies to avoid possible contaminations. Specifically we limit our sample 
only in galaxies with $i' < 24$, because the estimated photo-$z$ error is 
$\sigma_z = 0.02$ for galaxies with $i' <$ 24 and $z_{\rm ph} < 1.25$. 
Note that photo-$z$ errors for galaxies with $i'<25$ and $i'<25.5$ are
$\sigma_z = 0.04$ and 0.07, respectively.
Consequently we obtain 1553 galaxies with $1.17 < z < 1.20$ and $i' < 24$.

We find by cross-checking these two samples that 965 among 1553 photo-$z$ 
selected galaxies are \OII\ emitters selected by Takahashi et al (2007). 
The other 588 objects are galaxies without \OII\ emission (more strictly,
without \OII\ emission with EW $<12$ \AA\ rest-frame).
The SFR of \OII\ emitters are estimated through the relation of SFR $= 1.41
\times 10^{-41} L$([O{\sc ii}]) $M_\odot{\rm yr^{-1}}$ (Kennicutt 1998).
The minimum and maximum SFR values in our sample of the \OII\ emitters 
(with $i' < 24$) is 12 $M_\odot{\rm yr^{-1}}$ and 275 $M_\odot {\rm yr^{-1}}$, 
respectively.

The spatial distribution of all galaxies in our sample at $1.17 < z < 1.20$ is 
presented in Fig. 1. This figure shows that our sample includes galaxies in 
various density regions, and therefore the sample is suitable to investigate 
the environmental effects on the star-formation activity in galaxies at 
$z\simeq 1.2$.

\subsection{Contamination in the Photometric-Redshift Sample}

We estimate how our photo-$z$ sample is contaminated by foreground or background
objects. We use 22 objects with the spectroscopic redshifts 
(spec-$z$: class ``3'' and class ``4'' in Lilly et al. 2007) to examine the
accuracy of our photo-$z$ estimation.
Among these objects, 18 objects have spec-$z$ similar to our target range, 
$1.17 < z < 1.20$, within the photo-$z$ error, $\sigma_z=0.02$,  
while the remaining objects have spec-$z$ out of this redshift range.
The expected contamination rates for the photo-$z$ sample are thus estimated 
to be 18\%.
We also estimate the incompleteness of our photo-$z$ sample, using 22 objects 
with the spec-$z$ within our target range. Among these objects, 17 objects have
photo-$z$ in our target range within the photo-$z$ error. The expected 
incompleteness rate for the photo-$z$ sample are thus estimated to be 22\%.
From these results, we consider that at least about 80\% of photo-$z$ in our 
sample is reliable. We will show that this contamination and incompleteness do 
not affect on our results.

\section{RESULTS}
\subsection{Magnitude Dependence of the Emitter Fraction}
As described in \S 2.1, we selected 965 \OII\ emitters and 588 non-emitters. 
Accordingly, $\approx$ 62 \% of galaxies at $z\simeq 1.2$ are \OII\ emitters. 
We show the $i^\prime$-band number count of our sample in Fig. 2. This figure 
shows that the emitter fraction does not depend significantly on the 
$i^\prime$ magnitude. This indicates that more than half of galaxies at 
$z\simeq 1.2$ are active star-forming galaxies, regardless of galaxy magnitude.
This is completely different from the properties of \OII\ emitters in the local
universe. G\'omez et al. (2003) reported that the median \OII\ equivalent 
width of galaxies in the local universe in the Sloan Difital Sky Survey 
(SDSS; York et al. 2000) Early Data Release (EDR; Stoughton et al. 2002) sample
is $\sim 5$ \AA\, and the \OII\ emitter fraction 
(i.e., the rest-frame \OII\ equivalent width larger than 12 \AA\ ) is $\approx 
15$\%, much lower than our result at $z\simeq 1.2$ ($\approx 62$ \%). 
This is consistent to the redshift evolution of the SFRD between $z\simeq 1.2$
and $z\simeq 0$ (e.g., Lilly et al. 1996; Fujita et al. 2003).

\subsection{Environmental Effects on Star-Formation Activity}

Here we examine the relation between the star-formation activity and the galaxy
environment. We adopt the projected 10th-nearest-neighbor surface density as an 
indicator of the galaxy environment, following the manner in Dressler (1980). 
This density is estimated by using the distance to the 10th nearest neighbor 
from each object. This distance defines used to estimate the surface density as
\begin{equation}
\Sigma_{\rm 10th} = \frac{11}{\pi r^2}. 
\end{equation}
Galaxies near the edge of our field of view, i.e., the $r_{\rm10th}$ is greater
than the distance from the edge, are not used in the analysis since the local 
galaxy density of such galaxies is not correctly estimated. 
As discussed in \S 2.1, we limit our sample only in galaxies with $i' < 24$.
This limiting magnitude corresponds to $M(V)=-20.5$, adopting the typical 
SED in our sample. This limit is close to those used by G\'omez et al. (2003) 
[$M(r^*)=-20.6$] and Dressler (1980) [$M(V)=-19.7$]; note that these magnitudes
are re-calculated by using the cosmology adopted in our paper. 
Our local galaxy density of 1 ${\rm Mpc}^{-2}$ corresponds to $\approx 1.15$ 
${\rm Mpc}^{-2}$ if adopting the Cosmology used by Gomez et al. (2003). 
Note that Dressler (1980) calculated the local galaxy density differently from 
our method; i.e., we eliminate foreground and background galaxies using 
photometric redshift. Thus our 10th nearest neighbor is much farther away than 
that obtained through Dressler's definition. Therefore we do not compare our 
results with those of Dressler (1980).

To quantify the typical star-formation activity of galaxies as a function of 
the local galaxy density, we use the fraction of \OII\ emitters.
We expect a positive or negative dependence of the \OII\ emitter fraction on 
the number density of galaxies, if the star-formation activity at $z\sim 1$ 
has any preferences on galaxy environments.
We show the relation between the \OII\ emitter fraction and the local galaxy 
density in Fig. 3. 
We then apply a linear fit to obtain the relation,  
${\rm fraction} = (0.098\pm 0.081) \log \Sigma_{\rm 10th} + (0.611 \pm 0.023)$.
We find no significant correlation between the \OII\ 
emitter fraction and the local galaxy density.
The star-formation activity is almost constant at any local densities,
although we see a marginal signature of positive correlation between the 
star-formation activity and the local galaxy density.
This trend appears to be different from the relation in the local universe.
It should be noted that the redshift bin size in our sample ($\Delta z=0.03$) is 
comparable to the photo-$z$ error ($\sigma_z=0.02$). In order to examine how 
the photo-$z$ uncertainty affects our results, we have made a similar analysis 
adopting a larger redshift bin size of $\Delta z=0.07$, considering the photo-$z$ error. 
In this case, we still obtained the result that the star formation activity 
tends to be higher in higher density regions. To be consistent with previous 
studies (Dlessler 1980), we used a criterion of the 10th 
nearest neighbor. We have tried to adopt a lower value (3rd and 5th) and found 
the same trend of the results shown in Fig. 3.

\subsection{Effects of the Contamination and Incompleteness in Sample}

Here we consider the possible influence of the contamination and 
incompleteness on the photo-$z$ catalog. As mentioned in \S 2.2, the estimated 
photo-$z$ contamination and incompleteness is $\sim 20\%$. It is thought that our 
main result is not affected by the contamination and incompleteness, 
except for the case that the fraction of contamination or incompleteness 
depends strongly on the number density of galaxies. This is unlikely, 
because the photo-$z$ errors mainly depend on the photometric accuracy, 
which is independent on the 
environment except for highly confused regions (not majority in our sample).
Even if we overestimate or underestimate $\Sigma_{\rm 10th}$, 
the data points on Fig. 3 would shift just slightly along the 
$\Sigma_{\rm 10th}-{\rm axis}$.  
We thus conclude that the contamination and incompleteness of the photo-$z$ 
sample does not affect significantly on our results and discussion.

\subsection{Environmental Effects on the Galaxy Morphology}

Our results imply that star-formation activity is also high in high density 
regions at $z\simeq 1.2$, showing a different trend from that seen in the local
universe (Fig.3). Here we consider what is the origin of this difference 
between $z\sim 1$ and $z\sim 0$.
As a possible origin, we expect the influence of the galaxy interaction or 
merger because it is known that companion galaxies (galaxy interactions and 
mergers) can trigger star formation (e.g., Taniguchi \& Wada 1996).
We investigate environmental effects on the fraction of non-isolated galaxies
(non-isolated fraction) to study if galaxy interaction and mergers relate to 
star-formation in high density regions at $z\sim 1$.
To define the subsamples of isolated galaxies and non-isolated galaxies, 
we searched for the nearest neighbor for each galaxy in the sample satisfying 
the following two conditions:
\begin{equation}
1.17 < z_{\rm ph} < 1.20
\end{equation}
and
\begin{equation}
|i'_t-i'_n| < 2.5
\end{equation}
where $i'_t$ and $i'_n$ are the $i'$ magnitudes of a target object and its 
nearest neighbor, respectively. We define non-isolated galaxies as objects 
that have the nearest neighbors within the apparent separation less than 10 
arcsec. 
This corresponds to $\approx$ 80 kpc in proper distance, being similar to the 
distance between our Galaxy and its satellite galaxies. The remaining objects 
are defined as isolated galaxies. We note that our non-isolated galaxies does 
not include galaxies having a very small effect of the gravitational 
interaction between galaxies since we adopt the following condition: 
$|i'_t-i'_n|<2.5$.

We find that the fraction of the non-isolated galaxies is 37\% over the whole 
sample, including both the 43\% \OII\ emitters and 28\% non-emitters (Table 1). 
And, 71\% of all non-isolated galaxies in the sample are \OII\ emitters (Table 2), 
suggesting that interacting and/or merging galaxies tend to be star-forming 
galaxies. Based on the results of our classification, we study the relation 
between the non-isolated fraction among the \OII\ emitters and the local galaxy 
density.
Here we again adopt $\Sigma_{\rm 10th}$ as a measure of the local density. 
Fig. 4 shows the results of this relation. The linear fit gives the relation, 
${\rm fraction} = (0.159\pm 0.073) \log \Sigma_{\rm 10th} + (0.361\pm 0.019)$. 
This result shows that the non-isolated fraction is higher in denser regions
and suggest that galaxies in high density regions have a greater tendency to
interact or merge than low density regions.
In order to confirm whether interacting and merging galaxies are related to star-
formation in high density regions, we plot the non-isolated fraction among the 
\OII\ emitters as a function of the local galaxy density (Fig. 5).
Fig. 5 shows that the non-isolated fraction among the \OII\ emitters increases with
increasing local density. These results indicate that galaxy interactions 
and/or mergers tend to occur in higher density regions.  
We make the same analysis with $|i'_t-i'_n|<1.5$ as a test, and find the same
trend as seen in Fig. 4 and Fig. 5.

\section{The origin of the environmental effects}
In the local universe, it is well known that the star formation in galaxies 
is inactive in high density regions 
(e.g., Gomez et al. 2003; Mateus et al. 2004). 
For instance, Gomez et al. (2003) found an overall shift of \OII\ equivalent 
width distributions to lower values in 
high-density regions based on their analysis on the SDSS data.
However our results show that the relation between the fraction of \OII\ 
emitters and the local galaxy density at $z\simeq 1.2$ appears to be flat, 
or to show a slight positive correlation (Fig. 3). 
This implies that star formation in high density regions is also active, 
showing the different trend  from that seen in the local universe.

Recently, Elbaz et al. (2007) and Cooper et al. (2008) reported that the mean 
SFR of galaxies is higher in higher density regions at $z\sim 1$, that is completely 
different from the tendency seen at $z\sim 0$.
Specifically, Elbaz et al. (2007) studied the mean SFR of galaxies in the GOODS
fields at $z\sim 1$ by using both UV and IR luminosity, and found that it 
depends positively on the local galaxy density. 
On the other hand, using the SFR measured with \OII\ luminosities, 
Cooper et al. (2008) also studied that the mean SFR-local galaxy density 
relation in the DEEP2  and the SDSS and found a similar trend.
Both analysis indicate that the star-formation activity in high density 
regions is higher than that in low density regions at $z\sim 1$, unlike the 
case of the local universe. This gives close agreement with our results on 
the relationship between the \OII\ fraction and density that the environmental
dependence of the star-formation activity in galaxies is completely different
between $z\sim 1$ and $z\sim 0$.
 
Here we discuss the following question; what is the origin of this difference 
between $z\sim 1$ and $z\sim 0$? Such a different trend was predicted by 
semi-analytic models for the hierarchical galaxy formation, in the sense that
galaxies in the denser regions evolve more earlier in the history of universe 
(Elbaz et al. 2007).
Moreover, it is also claimed observationally that the evolution of galaxies 
occurred earlier in higher density regions, that is suggested by the discovery 
that the build-up of the color-magnitude relation appears to be delayed in 
lower-density environments (Tanaka et al. 2005).
As a possible origin of the difference between  $z\sim 1$ and $z\sim 0$, 
we expect the influence of the galaxy interaction or merger because
it is known that galaxy interaction and merger trigger intense star 
formation (e.g., Taniguchi \& Wada 1996; Mihos \& Hernquist 
1996; Taniguchi at al. 1999).  
That is, many galaxies in higher density regions could have experienced 
interaction 
and/or merging with star formation in the past, and then the star 
formation in them becomes inactive because they gradually consume their gas.
As a consequence, the number density of active star-forming galaxies could 
shift from high density regions to low density regions with time.

If our interpretation is correct, it is expected that there are many 
interacting and/or merging galaxies in higher density regions at high redshift.
In order to examine this expectation, 
we studied the morphology of the galaxies, as already described in \S 3.4.  
We found that the fraction of non-isolated galaxies in the \OII\ emitter 
sample is higher than in the non-\OII\ emitter sample (see Tables 1 and 2). 
In addition, the non-isolated fraction among the whole sample (including both 
\OII\ and non-\OII\ emitters) increases with increasing local galaxy density 
(Fig. 4) and the non-isolated fraction among the \OII\ emitters is higher in denser
region (Fig. 5). These results show that galaxy interactions and/or mergers tend to
occur in higher density regions, inducing their active star formation.
Note that the non-isolated fraction among the \OII\ emitters does not reach unity
in high density regions, although it is significantly higher than this in lower 
density regions. We speculate that some advanced-merger and/or minor-merger systems 
are included in the \OII\ emitter at high density regions, which seems to be 
hard to
be recognized as non-isolated galaxies morphologically. Our results are 
consistent with our expectation that interaction and/or merger induced 
star formation is a possible origin of star formation in high density regions 
at $z\simeq 1.2$.

Here we mention that the ram-pressure affects the star-formation activity
in high density regions in addition to the galaxy interaction and/or mergers.
This is because the gas stripping due to the ram-pressure generally quenches
the star formation in clusters of galaxies 
(e.g., Gunn \& Gott 1972; Hester 2006).
However, the ram-pressure does not significantly influence our results, 
since our sample 
does not include very high density regions such as central regions of clusters
of galaxies.
Therefore we consider that the star-formation activity of galaxies in high 
density regions is induced by the galaxy interaction, at $z\sim 1$.
Here we point out that the gas fraction in galaxies is higher at $z\sim 1$ than
that at $z\sim 0$ even in high density regions.
This difference results in that the galaxy mergers at $z\sim 1$ tend to be 
``wet'', while the galaxy mergers at $z\sim 0$ tend to be ``dry''
(e.g., Lin et al. 2008).
Since the wet mergers naturally result in active star formation 
(e.g., Woods et al. 2006; Bridge et al. 2007), the star formation is induced by
interacting and/or merging galaxies even in high density regions, at $z\sim 1$.
On the contrary, galaxy interactions and mergers in high density regions do not
cause active star formation at $z\sim 0$.
This picture is completely consistent to the results of Elbaz et al. (2007) 
and of Cooper et al. (2008). 

\section{SUMMARY}

In this paper, we present the relation between the star formation activity and 
the galaxy density at $z=1.17$--$1.20$, using 1553 objects with $i^{\prime} < 
24$ in the COSMOS field. We estimate star-formation activity by focusing on the
``emitter fraction''. Our main results are as follows.

(1) The relation between the fraction of \OII\ emitters and the local galaxy 
density at $z \simeq 1.2$ is flat, or positively correlated possibly, 
suggesting a different trend from the tendency observed in the local universe.

(2) The fraction of non-isolated galaxies (interactions, mergers, and 
multiples) increases with increasing density, and that fraction is 
systematically higher in \OII\ emitters than in non \OII\ emitters.

Through the observational studies in the local universe, 
it is known that star formation is inactive at high density regions. However, 
our results indicate that the star formation is active also in high density 
regions at $z\simeq 1.2$. We conclude that the origin of this difference is the
influence of ``wet'' galaxy interactions and/or mergers.
That is, galaxies in high density regions have experienced interacting and/or
merging with star formation at $z\sim1$, while the star formation in them
becomes because their gas is gradually consumed. And so the number
density of active star-forming galaxies shifted from high density regions to 
low density regions with time.
This implies that the evolution of galaxies in high density regions
occurs earlier, relative to that in low density regions. 

\acknowledgements

We would like to thank both the Subaru and HST staff for their invaluable help
and to acknowledge the anonymous referee for an extremely useful report.
This work was financially supported in part by the Ministry
of Education, Culture, Sports, Science, and Technology (Nos. 10044052 
and 10304013), and by JSPS (15340059, 17253001, and 19340046). 


\begin{figure}
\epsscale{1.0}

\includegraphics[scale=1.0, angle=270]{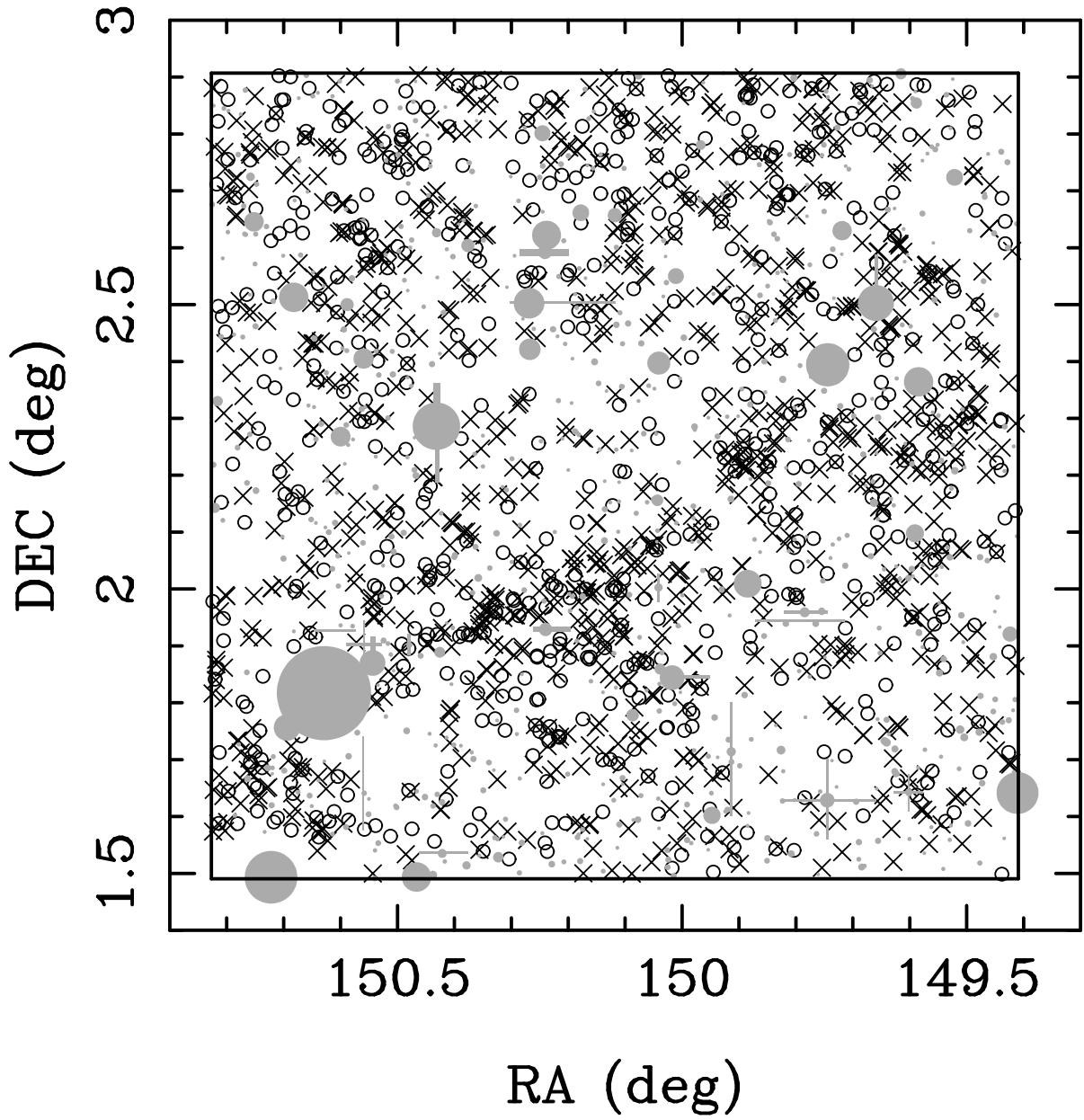}
\caption{Spatial distributions of our sample at $1.17 < z < 1.20$ in the 
COSMOS field. Open circles denote non-emitters in our sample. 
Cross symbols denote \OII\ emitters selected by Takahashi et al. (2007) 
in our sample. The circle areas are masked out for the detection.
We select galaxies in the region of the large solid square: 
$149^{\circ}.40917 <$ RA (J2000)
$< 150^{\circ}.82680$ and $1^{\circ}.49056 <$ Dec (J2000)$< 2^{\circ}.90705$ 
(Takahashi et al. 2007).}
\end{figure}

\begin{figure}
\epsscale{1.0}

\includegraphics[scale=0.7, angle=270]{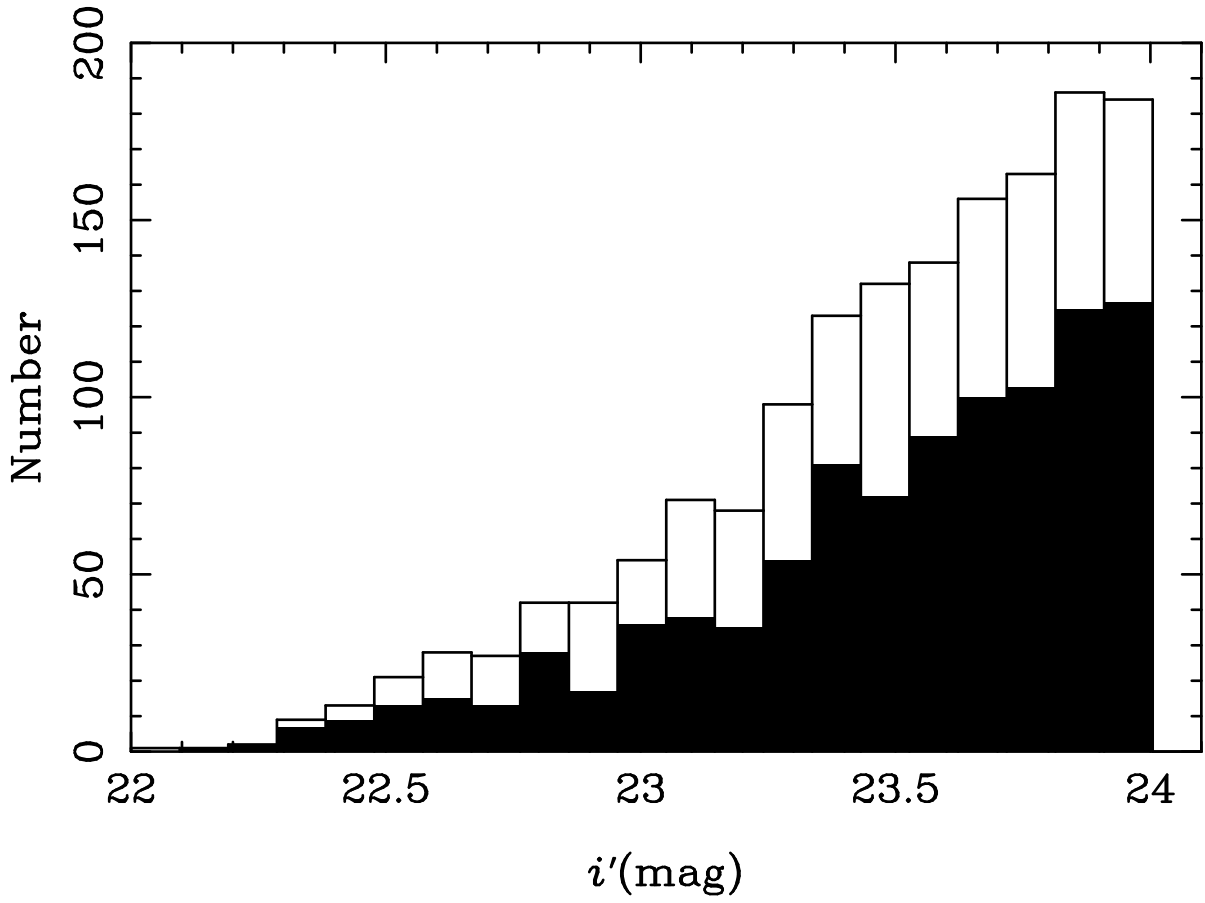}

\caption{
The number count of our sample as a function of $i^{\prime}$ magnitude. 
Open boxes denote our total sample, including both \OII\ emitters and non 
emitteres.  Filled boxes denote \OII\ emitters. 
}
\end{figure}

\begin{figure}
\epsscale{1.0}
\plotone{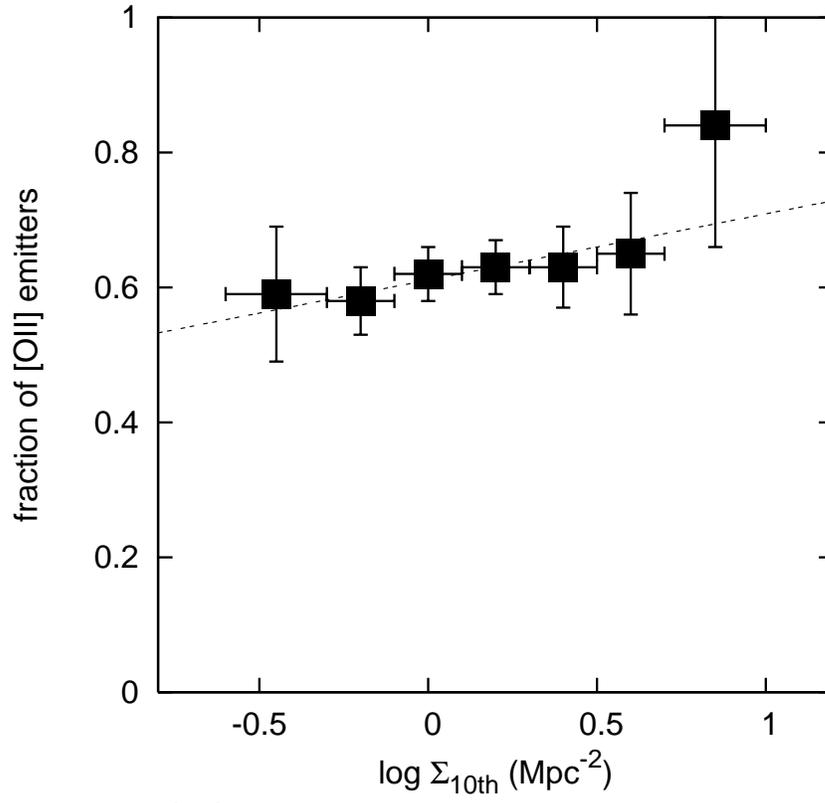}
\caption{
The fraction of \OII\
emitters is plotted as a function of the galaxy local density. 
The dotted line shows the linear fit to the data. 
The star-formation activity is more active in higher density regions.
Here the adopted size of the log $\Sigma_{\rm 10th}$ bin is 0.2 dex, 
except for the lowest- and highest-log$\Sigma_{\rm 10th}$ bins where
there are only a few objects (0.3 dex for these bins). 
}
\end{figure}


\clearpage

\begin{deluxetable}{lcc}
\tablecaption{\label{La:tab:CV}The fraction of isolated galaxies and non-isolated galaxies.}
\tablewidth{0pt}
\tablehead{
\colhead{} & 
\colhead{\OII\ emitters} & 
\colhead{non-emitters}
}
\startdata
isolated fraction  & $57\%$ & $72\%$ \\
non-isolated fraction & $43\%$ & $28\%$\\
\hline
total & $100\%$ & $100\%$\\
\enddata                                                                       
\end{deluxetable}


\begin{deluxetable}{lcc}
\tablecaption{\label{La:tab:CV} The fraction of \OII\ emitters and 
non-emitters.}
\tablewidth{0pt}
\tablehead{
\colhead{} & 
\colhead{isolated galaxies} & 
\colhead{non-isolated galaxies}
}
\startdata
\OII\ fraction  & $56\%$ & $71\%$ \\
non-emitter fraction & $44\%$ & $29\%$\\
\hline
total & $100\%$ & $100\%$\\
\enddata                                                                       
\end{deluxetable}
\clearpage
\begin{figure}
\epsscale{1.0}
\plotone{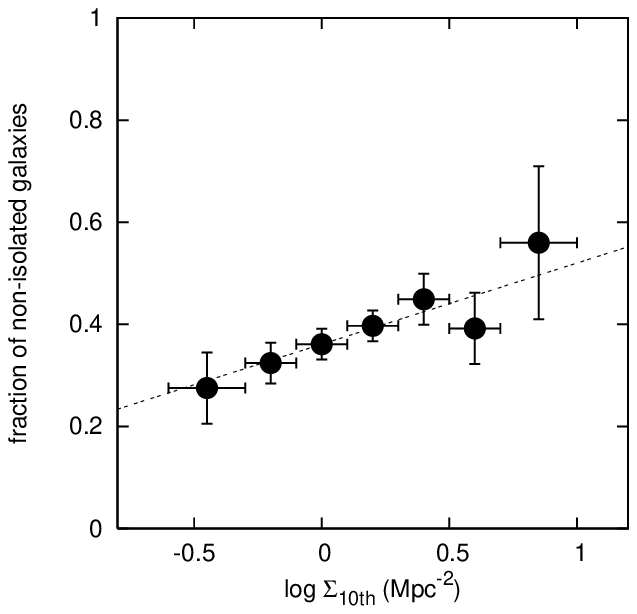}
\caption{
The fraction of non-isolated galaxies is plotted as a function of the 
galaxy local density. The dotted line shows the linear fit to the data.
Here the adopted size of the log $\Sigma_{\rm 10th}$ bin is 0.2 dex, 
except for the lowest- and highest-log$\Sigma_{\rm 10th}$ bins where
there are only a few objects (0.3 dex for these bins).
}
\end{figure}

\begin{figure}
\epsscale{1.0}
\plotone{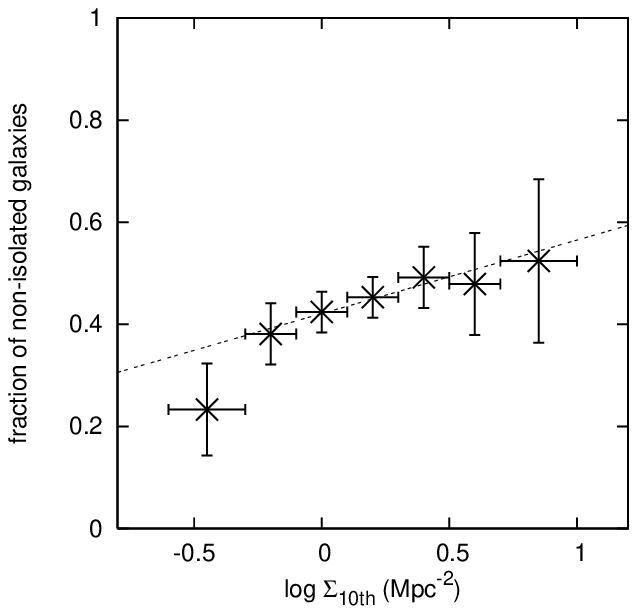}
\caption{
The fraction of non-isolated galaxies among \OII\ emitters is plotted 
as a function of the galaxy local density. The dotted line shows the 
linear fit to the data.
Here the adopted size of the log $\Sigma_{\rm 10th}$ bin is 0.2 dex, 
except for the lowest- and highest-log$\Sigma_{\rm 10th}$ bins where
there are only a few objects (0.3 dex for these bins).
}
\end{figure}

\end{document}